\documentclass[10pt,a4paper]{article}

\usepackage[utf8]{inputenc}
\usepackage[T1]{fontenc}
\usepackage{comment,xcolor} % For editing 
\usepackage[page,toc,title]{appendix}
\usepackage{amsmath,amssymb,amsfonts} % Math symbols
\usepackage{graphicx} % Including figures
\usepackage{booktabs,caption, siunitx} % For tables
\usepackage[margin=1in]{geometry} % Page margins
\usepackage{natbib} % For bibliography management
\usepackage{booktabs} % For professional tables
\usepackage{hyperref} % For clickable links and references
\hypersetup{
    colorlinks=true,
    linkcolor=blue,
    filecolor=magenta,      
    urlcolor=cyan,
    citecolor=blue,
}
\usepackage{caption} % For captions
\usepackage{enumitem} % For customized lists (optional)
\usepackage{sectsty} % To change section heading styles (optional)
\usepackage{times} % Using Times font (common for publications)

\makeatletter
\def\keywords{\@ifnextchar\bgroup{\@keywords}{\@exkeywords}}
\def\@keywords#1{\par\addvspace\baselineskip\noindent\keywordname\enspace\ignorespaces#1}
\def\@exkeywords{\par\addvspace\baselineskip\noindent\keywordname\enspace}
\gdef\keywordname{\textbf{Keywords:}}
\def\subjclass{\@ifnextchar\bgroup{\@subjclass}{\@exsubjclass}}
\def\@subjclass#1{\par\addvspace\baselineskip\noindent\subjclassname\enspace\ignorespaces#1}
\def\@exsubjclass{\par\addvspace\baselineskip\noindent\subjclassname\enspace}
\gdef\subjclassname{\textbf{Mathematics Subject Classification (MSC 2020):}}
\makeatother

\title{Modular versus Hierarchical: A Structural Signature of Topic Popularity in Mathematical Research}
\author{
    Brian Hepler\thanks{ORCID: 0000-0002-8037-930X. The full source code, data, and reproducible analysis pipeline for this study are publicly available in our GitHub repository: \url{https://github.com/brian-hepler-phd/MRC-Network-Analysis}} \\
    \texttt{hepler.brian@gmail.com} \\
}
\begin{document}

\maketitle

\begin{abstract}
Mathematical researchers, especially those in early-career positions, face critical decisions about topic specialization with limited information about the collaborative environments of different research areas. The aim of this paper is to study how the popularity of a research topic is associated with the structure of that topic's collaboration network, as observed by a suite of measures capturing organizational structure at several scales. We apply these measures to 1,938 algorithmically discovered topics across 121,391 papers sourced from arXiv metadata during the period 2020--2025. Our analysis, which controls for the confounding effects of network size, reveals a structural dichotomy--we find that popular topics organize into modular ``schools of thought,'' while niche topics maintain hierarchical core-periphery structures centered around established experts. This divide is not an artifact of scale, but represents a size-independent structural pattern correlated with popularity. We also document a ``constraint reversal'': after controlling for size, researchers in popular fields face greater structural constraints on collaboration opportunities, contrary to conventional expectations. Our findings suggest that topic selection is an implicit choice between two fundamentally different collaborative environments, each with distinct implications for a researcher's career. To make these structural patterns transparent to the research community, we developed the Math Research Compass (\url{https://mathresearchcompass.com}), an interactive platform providing data on topic popularity and collaboration patterns across mathematical topics.

\keywords{scientific collaboration, network science, topic modeling, author name disambiguation, collaboration networks}

\subjclass{01A80, 91D30, 05C82, 62R07} % MSC codes
\end{abstract}

\section{Introduction}
The structure of scientific collaboration networks shapes knowledge flow and the professional environments in which researchers build their careers. In mathematics, where collaboration patterns are highly varied, understanding these structural differences is important for both individual career navigation and institutional science policy \citep{Wuchty2007}.

The evolution of such networks is governed by two distinct classes of mechanisms. Endogenous mechanisms are processes internal to the network's topology, such as preferential attachment, where a node's existing connectivity drives the formation of new links \citep{Barabasi1999}. In contrast, exogenous mechanisms are driven by factors external to the network structure, such as homophily based on shared attributes like research interests or institutional affiliation \citep{McPherson2001, Khanam2022}. While network formation theory distinguishes between these processes \citep{jackson1996strategic}, empirically separating their effects in observational data is a well-known identification problem \citep{manski1993identification, shalizi2011homophily, tomasello2014role}. It is therefore challenging to determine whether an observed network property is a genuine structural pattern associated with an external factor or simply a predictable artifact of network size and growth.

This problem is particularly relevant in mathematics, where research is organized into numerous subfields with distinct collaborative norms \citep{Newman2001, brunson2014evolutionary}. The popularity of a research area, a proxy for community size and collective attention, is a key exogenous variable likely to be associated with network structure. However, its influence is difficult to disentangle from the endogenous effects of network scale.

This paper addresses this challenge by applying a two-stage statistical analysis to the collaboration networks of 1,938 algorithmically-identified mathematical topics. We first perform a baseline comparison of network metrics between popular and niche topics. Second, we use regression models to control for network size, allowing us to distinguish size-dependent effects from size-independent structural patterns associated with popularity. This approach allows us to test which network properties are robustly associated with the exogenous popularity of a field, versus those that are better explained as endogenous scaling effects. While popularity itself may be influenced by network properties over long time scales, we treat it as a given characteristic of research areas for the purposes of this cross-sectional analysis.

Our analysis reveals a robust structural dichotomy: popular topics organize into modular "schools of thought," while niche topics maintain hierarchical core-periphery structures. We also document a "constraint reversal," where, contrary to initial expectations, researchers in popular fields face greater structural constraints after controlling for size. This approach identifies which structural patterns warrant further investigation through longitudinal or experimental methods to establish causal mechanisms.

This paper proceeds as follows. Section \ref{sec:methods} details our methodology for topic identification, network construction, and the two-stage statistical analysis. Section \ref{sec:results} presents our empirical findings, documenting both the initial structural differences and the results after controlling for network size. Section \ref{sec:discussion} discusses implications for understanding mathematical collaboration environments and their relevance to career navigation. We conclude by acknowledging limitations and suggesting directions for future research linking network structures to career outcomes.

\section{Methods}\label{sec:methods}

\subsection{Dataset and Topic Identification}

We analyzed the Cornell ArXiv dataset (\cite{Clement2019}), which contains metadata for approximately 2.7 million scientific papers. Our analysis focused on mathematics papers published between 2020 and 2025, yielding 121,391 papers across 31 mathematical subfields as classified by the ArXiv Mathematics Subject Classification system.

Research topics were identified using BERTopic \citep{Grootendorst2022}, a state-of-the-art topic modeling approach. The BERTopic pipeline consists of three main stages. First, it generates semantic vector embeddings for each document using a Bidirectional Encoder Representations from Transformers (BERT) model \citep{devlin2019}. Next, it applies Uniform Manifold Approximation and Projection (UMAP) to these embeddings to reduce their dimensionality while preserving local and global structure \citep{McInnes2018}. Finally, it uses Hierarchical Density-Based Spatial Clustering of Applications with Noise (HDBSCAN) on the dimension-reduced embeddings to identify dense clusters of documents, which constitute the research topics \citep{McInnes2017}. We concatenated paper titles and abstracts as input text, applied preprocessing to remove common stopwords and mathematical notation artifacts, and generated semantic embeddings using the \texttt{all-MiniLM-L6-v2} sentence transformer model. UMAP reduced embeddings to 5 dimensions using 15 neighbors and cosine distance, followed by HDBSCAN clustering with a minimum cluster size of 15 papers. This process identified 1,938 distinct research topics, with each paper assigned to its most coherent topic based on the highest probability score. Each topic was assigned a primary mathematical category based on the most frequent ArXiv category among its constituent papers.

\subsection{Author Name Disambiguation}
A prerequisite for constructing accurate collaboration networks is robust author name disambiguation (AND). Given that raw author strings from bibliographic sources often contain variations, initials, and ambiguities that can lead to either incorrectly splitting single authors into multiple nodes or incorrectly merging distinct authors into one, we implemented a dedicated multi-stage AND pipeline \texttt{AuthorDisambiguator}; see Appendix \ref{sec:AND_pipeline} for full algorithm details.

Our multi-stage process aligns with current best practices in the field, which have converged on sequential pipelines that combine graph-based similarity with network-aware clustering to achieve high precision on large-scale datasets (\cite{Rehs2021}; \cite{Chen2023}). We adopted a particularly conservative approach, prioritizing high precision to minimize the introduction of false connections that could artificially distort network topology (\cite{Kim2019}; \cite{Strotmann2009}).

The pipeline began by parsing and normalizing all unique author name strings ($N \approx 120,000$) from the 121,391 papers. Normalization involved lowercasing, diacritic removal, and standardization of particles and punctuation. Subsequently, a graph-based approach was employed:
\begin{enumerate}
\item \textbf{Initial Similarity Graph Construction}: An author similarity graph was constructed where nodes represent unique normalized names. Edges were added between names sharing the same last name if they met strict string similarity criteria (e.g., initial expansions like `J. Doe' to `John Doe' or high Levenshtein distance > 0.95).

\item \textbf{Cluster Merging}: Connected components in this graph (clusters of potentially identical authors) were analyzed. A cluster was merged into a single canonical author profile if all full first names within it were determined to be compatible variants (e.g., `Robert', `Rob'). Specific heuristics were applied to handle potentially ambiguous patterns, such as those common in Romanized East Asian names, by employing stricter similarity thresholds for first name compatibility. Clusters exceeding a size threshold (50 names) or containing multiple clearly distinct first names (e.g., `Pengcheng Xie' and `Pengxu Xie' within the same last-name block) were handled conservatively, with the algorithm only attempting to identify and merge highly confident sub-groups.

\item \textbf{Network-Based Co-occurrence Merging}: Remaining canonical profiles were further compared based on co-authorship patterns and shared publications. Pairs of authors sharing the same last name and first initial were evaluated using a weighted Jaccard similarity of their co-author sets and paper sets. Pairs exceeding a similarity threshold (0.5) and minimum paper count (2) were merged.

\end{enumerate}
This iterative process reduced the initial ~120,767 unique raw author strings to 117,883 canonical author profiles, representing a 2.39\% reduction. Manual validation of 100 randomly sampled merges performed by the pipeline confirmed a precision of 86.0\%. While this represents a substantial improvement over simple standardization, residual disambiguation errors may remain and are acknowledged as a limitation (see Section \ref{subsec:strengths_limitations}).

\subsection{Collaboration Network Construction}

For each of the 1,938 research topics, we constructed an undirected co-authorship network using the disambiguated canonical author profiles generated by our AND pipeline. In these networks, nodes represent unique authors active in the topic, and edges represent a co-authorship relationship on at least one paper within that topic.

The construction process followed established protocols for creating collaboration networks from bibliographic data (\cite{Newman2001}). For each paper with multiple authors, a complete graph (or clique) was formed among all co-authors. This means an edge was created between every pair of authors on a given paper. 
\begin{description}
    \item[\textbf{Edge Weights:}] The networks are weighted. If a pair of authors co-authored multiple papers within the same topic, the weight of the edge between them was incremented for each joint publication. These weights represent the strength of the collaborative tie and were used in weighted network metric calculations, such as modularity.
    \item[\textbf{Node Attributes:}] Each node (author) was attributed with the total count of papers they published within that specific topic.
    \item[\textbf{Self-Loops:}] Self-loops were prevented by ensuring the author list for each paper contained unique individuals before edge creation.
\end{description}
Only papers with multiple authors contributed to the formation of network edges. However, papers with a single author were retained in our dataset for the calculation of topic-level metrics like the overall collaboration rate.

\subsection{Network Metrics}

To characterize the structural properties of mathematical collaboration networks, we employ a carefully selected set of 10 network metrics organized into four domains. This captures distinct yet complementary aspects of network structure, providing a comprehensive structural signature of each mathematical subfield. 

\begin{table*}[htbp]
  \centering
  \caption{A Multi-Scale Framework of Network Metrics for Collaboration Analysis}
  \label{tab:network_metrics}
  \resizebox{\textwidth}{!}{%
  \begin{tabular}{@{}lll@{}}
    \toprule
    \textbf{Analysis Domain} & \textbf{Metric Name} & \textbf{Description} \\
    \midrule
    \textbf{Collaboration Dynamics} & Collaboration Rate & Proportion of multi-author papers, indicating overall teamwork propensity. \\
    \textit{(Tie Strength \& Persistence)} & Repeated Collaboration Rate & Proportion of author pairs with multiple joint papers, measuring stability of research teams. \\
    \addlinespace
    
    \textbf{Global Topology} & Degree Centralization & Concentration of collaborations around a few central "hub" researchers (hierarchy). \\
    \textit{(Overall Network Shape)} & Degree Assortativity & Tendency for highly-collaborative authors to work with other high-prolific authors. \\
                              & Small-world Coefficient ($\omega$) & Balance of high local clustering with short global path lengths for efficient diffusion. \\
                              & Robustness Ratio & Network's resilience to the targeted removal of hubs versus random failures. \\
    \addlinespace

    \textbf{Mesoscopic Organization} & Modularity (Q) & Strength of a network's division into distinct communities or "schools of thought". \\
    \textit{(Community \& Core Structure)} & Coreness Ratio & Proportion of authors belonging to the network's densely connected core. \\
    \addlinespace

    \textbf{Researcher Positioning} & Average Constraint & Extent to which an individual's connections are redundant (inverse of brokerage). \\
    \textit{(Individual-level Opportunity)} & Average Effective Size & Average number of non-redundant connections per researcher, measuring access to diverse information. \\
    \bottomrule
  \end{tabular}%
  }
\end{table*}

\begin{comment}
\begin{table}[htbp]
  \centering
  \caption{Network Metrics Used for Collaboration Structure Analysis}
  \label{tab:network_metrics}
  \begin{tabular}{@{}lll@{}}
    \toprule
    \textbf{Category} & \textbf{Metric Name} & \textbf{Description} \\
    \midrule
    \textbf{Collaboration Dynamics} & Collaboration Rate & Proportion of multi-author papers \\
                                   & Repeated Collab. Rate & Proportion of author pairs with multiple joint papers \\
    \addlinespace
    \textbf{Network Topology} & Degree Centralization & Concentration of connections around hub nodes \\
                              & Degree Assortativity & Tendency for similar-degree nodes to connect \\
                              & Modularity & Strength of community structure via Louvain algorithm \\
                              & Small-world Coefficient & Ratio of clustering to path length relative to random networks \\
    \addlinespace
    \textbf{Structural Resilience} & Coreness Ratio & Proportion of authors in the maximum k-core \\
                                   & Robustness Ratio & Resilience to targeted vs. random node removal \\
    \addlinespace
    \textbf{Researcher Positioning} & Average Constraint & Burt's measure of structural holes \\
                                    & Average Effective Size & Number of non-redundant connections per researcher \\
    \bottomrule
  \end{tabular}
\end{table}
\end{comment}

\subsubsection{Collaboration Dynamics (Measuring Tie Strength and Persistence)}

This domain quantifies the intensity and persistence of collaborative relationships. Following the theory of tie strength by \citet{granovetter1973strength} and its recent experimental validation \citep{Rajkumar2022}, we measure:

\textbf{Collaboration rate}, the proportion of papers with multiple authors, provides a baseline measure of a field's propensity for teamwork. This metric captures the overall collaborative culture within mathematical subfields, following the foundational approach of \citet{Newman2001}.

\textbf{Repeated collaboration rate}, the fraction of collaborations that occur more than once between the same pair of researchers. Building on the operationalization of collaboration persistence by \citet{Payumo2021}, this metric distinguishes between fields characterized by transient partnerships and those fostering stable research teams. The work of \citet{Petersen2015} on ``super ties" demonstrates that such persistent collaborations create multiplicative effects on scientific productivity.

Together, these metrics reveal whether mathematical subfields operate through short-lived project-based teams or cultivate long-term collaborative relationships, with direct implications for knowledge accumulation and research continuity.

\subsubsection{Global Network Topology (Characterizing Overall Network Shape and Resilience)}

This domain captures macro-level structural properties that determine network efficiency and resilience. We employ four complementary metrics:

\textbf{Degree centralization}, following the framework of \citet{Freeman1978} as applied to scientific networks by \citet{Newman2001}, measures the extent to which collaborations concentrate around a few highly connected ``hub" researchers. High centralization indicates hierarchical structures, while low centralization suggests more egalitarian patterns.

\textbf{Degree assortativity} measures whether highly collaborative researchers preferentially work with other highly prolific collaborators (assortative) \citep{Newman2002}. As \citet{Khanna2022} demonstrate, this choice involves a critical trade-off: assortative structures tend to produce higher research output but lower novelty, while disassortative structures generate more innovative outcomes.

\textbf{Small-world coefficient ($\omega$)}, calculated as $\omega = (C/C_{\text{random}})/(L/L_{\text{random}})$ \citep{watts1998}, captures whether networks optimize both high local clustering for specialized knowledge development and short global path lengths for efficient knowledge diffusion-a balance \citet{Newman2001} identified as characteristic of productive scientific fields.

\textbf{Robustness ratio} quantifies the network's structural resilience by comparing its tolerance to the targeted removal of high-degree ``hub" authors versus random node failures. A low ratio indicates a fragile, scale-free structure, providing crucial context for the stability of the collaborative landscape.

\subsubsection{Mesoscopic Organizational Structure (Identifying Community and Core-Periphery Patterns)}

This domain reveals intermediate-scale organization, reflecting how research fields self-organize into communities and hierarchies:

\textbf{Modularity (Q)}, using the modularity optimization approach of \citet{newman2006modularity}, quantifies the strength of a network's division into research communities or ``schools of thought." This is essential for understanding whether a field fragments into isolated clusters or maintains integrated collaborations.

\textbf{Coreness ratio}, building on the core-periphery framework of \citet{Borgatti2000} and advances by \citet{zhang2015}, measures the proportion of researchers belonging to the network's densely connected core. This reveals whether fields organize around expert-led hierarchies (high coreness) or distribute expertise more evenly.

\subsubsection{Nodal Positioning and Brokerage (Quantifying Individual-level Opportunity)}

This domain measures the average opportunity for individual researchers to access diverse information, based on the theory of structural holes \citep{Burt1992, Burt2004}:

\textbf{Average constraint} measures the extent to which an individual's network is redundant. Following Burt's formulation and recent clarifications \citep{everett2020unpacking}, lower constraint indicates greater opportunity to bridge structural holes, which is strongly linked to higher productivity and innovation in scientific collaboration \citep{yang2024higher, liao2024factors}.

\textbf{Average effective size} directly measures an individual's brokerage potential by quantifying their number of non-redundant contacts. Higher values indicate that the typical researcher has access to diverse information sources—what \citet{Burt2004} termed ``vision of options otherwise unseen."

\subsubsection{Integrated Framework Justification}

These ten metrics, spanning four complementary domains, provide a comprehensive structural signature that captures how mathematical research organizes across multiple scales. This multi-domain approach addresses limitations of single-metric analyses by capturing network properties at multiple scales--from individual positioning to global topology. By combining metrics of tie strength, macro-structure, meso-organization, and micro-positioning, we can identify distinct ``collaboration phenotypes" that characterize different mathematical subfields and explain variations in their productivity, innovation, and knowledge diffusion patterns. See Appendix \ref{sec:appendix_metrics} for more computational details regarding these metrics.

\subsection{Popular vs. Niche Topic Classification}

Topics were classified by research popularity based on total paper count. We rank-ordered all 1,938 topics by the number of papers and designated the top 20\% (387 topics, 74--2,207 papers each) as "popular" and bottom 20\% (387 topics, 11--17 papers each) as "niche." This symmetric approach ensured balanced sample sizes while capturing extreme differences in research attention.

The 20\% threshold was selected to identify topics with meaningfully different collaboration environments while maintaining adequate sample sizes for statistical analysis. To assess the robustness of our findings to this classification choice, we conducted two further checks. First, sensitivity analyses using 15\%, 20\%, 25\%, and 30\% cutoffs showed consistent patterns across thresholds. Second, we performed a regression analysis comparing our binary classification against multiple continuous measures of popularity (e.g., log-transformed paper count, rank percentile). For a clear majority of metrics (8 of 10), particularly those capturing network organization, the binary classification provided a superior model fit. For instance, the binary model for modularity (\text{Adj.~}$R^2 = 0.496$) substantially outperformed the best continuous model (\text{Adj.~}$R^2 = 0.399$), as did the model for coreness ratio (\text{Adj.~}$R^2 = 0.636$ vs. $0.554$). This finding suggests the existence of a genuine threshold effect, where collaborative dynamics undergo a qualitative shift once a topic achieves a critical level of popularity. The exceptions to this pattern were themselves informative: for metrics like collaboration rate, a continuous model was clearly superior (\text{Adj.~}$R^2 = 0.974$ vs. $0.335$), indicating these properties are likely governed by more universal, size-dependent scaling phenomena. This confirms that our central conclusions are not an artifact of the specific 20\% threshold but reflect a robust underlying phenomenon, with the binary approach best capturing the categorical shifts in organizational structure central to our investigation. Appendix \ref{sec:appendix_tables} (Table \ref{tab:supp_binary_vs_continuous}) shows detailed results from our analysis comparing binary and continuous popularity measures across all network metrics.

\subsection{Statistical Analysis}

Our analysis proceeded in two stages to disentangle popularity effects from network size effects:

\textbf{Stage 1: Baseline Comparisons.}
We first assessed distributional assumptions using Shapiro-Wilk tests for normality. Since all metrics violated normality assumptions (all $p < 0.001$), we employed non-parametric statistical tests throughout. This non-normality is expected in network data due to bounded metrics (e.g., collaboration rates constrained to 0-1), power-law degree distributions common in academic networks, and heterogeneous collaboration structures across scientific disciplines (\cite{Newman2004}, \cite{Newman2005}, \cite{Martini2021}).

Group comparisons used Mann-Whitney U tests to assess statistical significance. This non-parametric test operates by ranking all observations from both groups and calculating the U statistic, which is the number of times a value from one group is larger than a value from the other. This statistic is then used to compute a p-value against the null hypothesis that the two distributions are identical. This approach is robust to non-normal distributions and outliers common in network data.

To quantify the magnitudes of these differences, we calculated the non-parametric effect size Cliff's delta ($\delta$) (\cite{Cliff1993}), which estimates the degree of separation between two distributions as the difference in dominance probabilities (i.e., $\delta = P(X > Y) - P(Y > X)$). For interpretation, we use standard thresholds of 0.147 (small), 0.33 (medium), and 0.474 (large), which represent the non-parameteric equivalents of Cohen's original effect size benchmarks (\cite{Romano2006}).

A Bonferroni correction was applied for the ten comparisons, setting our significance threshold at $\alpha = 0.005$. Confidence intervals for key estimates were computed using bootstrap resampling with 10,000 iterations.

\textbf{Stage 2: Controlling for Network Size.} To test whether observed differences persisted beyond size effects, we conducted multiple regression analyses. For each network metric, we fit three models:
\begin{enumerate}
    \item \textbf{Simple Model:} Metric $\sim$ Popularity (binary)
    \item \textbf{Size Control Model:} Metric $\sim$ Popularity + log(Network Size)
    \item \textbf{Interaction Model:} Metric $\sim$ Popularity $\times$ log(Network Size)
\end{enumerate}

All variables were standardized (mean=0, SD=1) before analysis to ensure comparability of coefficients across metrics. A significant popularity coefficient in the size control model indicates a robust effect that persists beyond network scale. We also examined correlations between network size and each metric to assess confounding strength.

All analyses were performed in Python 3.11 using SciPy 1.15.2 and statsmodels 0.14.0.

\subsection{Robustness of Topic Modeling and Classification}

To ensure our findings were not artifacts of specific parameter choices, we conducted two comprehensive validation analyses for our topic modeling and classification approach.

First, to validate the BERTopic model itself, we performed a sensitivity analysis testing 72 different hyperparameter combinations on a 10,000-document subsample. This revealed that while the exact number of topics generated was moderately sensitive to the \verb+min_topic_size+ parameter (Coefficient of Variation = 0.44), the overall topic structure was reasonably stable (mean Adjusted Rand Index = 0.34 $\pm$ 0.06). This analysis informed our choice of baseline parameters for the main study (\verb+min_topic_size=15+, \verb+n_neighbors=15+, \verb+n_components=5+). Furthermore, to ensure the semantic coherence of the generated topics, we conducted a manual validation of a stratified sample of 100 topics. Appendix \ref{sec:appendix_tables} (Table \ref{tab:manual_BERT}) presents representative examples from this validation, demonstrating that BERTopic successfully identified genuine mathematical research areas across all size ranges. The full validation found that 74\% of topics were of high quality (scores 4-5) and 26\% were of medium quality (score 3), with no low-quality topics or algorithmic artifacts detected. Notably, even the smallest topics (15-17 papers) represented coherent specialized research areas such as causal fermion systems or Cherednik algebras, confirming that our minimum cluster size of 15 was appropriate.

Second, and most crucially, we generated three additional complete topic models using alternative minimum topic size values (10, 20, and 25), resulting in four distinct topic sets ranging from 1,164 to 2,953 topics. We then re-ran our entire two-stage statistical analysis—both the baseline Mann-Whitney U tests and the full regression models—on all four topic sets (see Appendix \ref{sec:appendix_tables}, Table \ref{tab:robustness_effect_sizes}). The central finding of our paper, the duality between scale-driven and popularity-driven network effects, remained consistent across all four independent analyses. The same metrics were robustly associated with popularity, and the same metrics were confounded by size in every single run (see Appendix \ref{sec:appendix_tables}, Table \ref{tab:appendix_interaction} for full interaction model results across one run). This demonstrates that our conclusions reflect a fundamental phenomenon in the data, independent of the specific granularity of the topic model.

\section{Results}\label{sec:results}

\subsection{Dataset Overview}

Our analysis encompassed 121,391 mathematics papers from which BERTopic identified 1,938 distinct research topics. The classification yielded 387 popular topics (mean = 314.2 $\pm$ 345.1 papers) and 387 niche topics (mean = 13.8 $\pm$ 1.7 papers), representing a 22.8-fold difference in mean paper counts. This extreme difference ensured distinct collaboration environments for comparison while maintaining balanced sample sizes.

\subsection{Baseline Differences Between Popular and Niche Topics}

Mann-Whitney U tests revealed significant differences in 9 of 10 network metrics between popular and niche topics (Table 1). All significant results survived Bonferroni correction ($\alpha = 0.005$), with most metrics showing large effect sizes by conventional standards for Cliff's delta.

Popular topics demonstrated substantially higher modularity (87.9\% vs. 68.8\%, $\delta = 0.79$), indicating stronger community structure and subdivision into distinct research groups. These networks also exhibited pronounced small-world properties, with coefficients nearly six-fold higher than niche topics (84.1 vs. 14.9, $\delta = 0.92$). Researchers in popular topics maintained larger effective network sizes (1.48 vs. 1.19, $\delta = 0.70$), suggesting more diverse, non-redundant connections. Conversely, these networks showed lower degree centralization (6.2\% vs. 14.2\%, $\delta = -0.56$) and dramatically reduced robustness to targeted node removal (32.9\% vs. 76.9\%, $\delta = -0.83$).

Niche topics exhibited contrasting organizational patterns characterized by hierarchical structures. These networks demonstrated exceptionally high coreness ratios (35.3\% vs. 7.0\%, $\delta = -0.93$), indicating strong core-periphery organization where established experts form tightly connected cores. Degree assortativity was also significantly higher (49.6\% vs. 29.3\%, $\delta = -0.34$), suggesting preferential connections between similarly connected researchers. Additionally, researchers in niche topics experienced greater structural constraint (91.8\% vs. 82.7\%, $\delta = -0.56$), reflecting more limited brokerage opportunities.

Only collaboration rate showed no significant difference between groups (75.5\% vs. 74.7\%, $p = 0.51$, $\delta = -0.03$), indicating that the propensity for multi-author publication is independent of topic popularity.

\begin{table}[htbp]
  \centering
  \caption{Initial Comparison of Network Metrics Between Popular and Niche Topics}
  \label{tab:initial_comparison}
  \small
  \begin{tabular}{@{}lccccl@{}}
    \toprule
    Metric & Popular Mean (SD) & Niche Mean (SD) & p-value & Cliff's $\delta$ & Effect Size \\
    \midrule
    Collaboration Rate & 0.755 (0.148) & 0.747 (0.196) & 0.511 & -0.027 & Negligible \\
    Repeated Collab. Rate & 0.148 (0.058) & 0.130 (0.143) & $<0.001$ & 0.288 & Small \\
    Degree Centralization & 0.062 (0.043) & 0.142 (0.111) & $<0.001$ & -0.559 & Large \\
    Degree Assortativity & 0.293 (0.226) & 0.496 (0.409) & $<0.001$ & -0.337 & Medium \\
    Modularity & 0.879 (0.081) & 0.688 (0.173) & $<0.001$ & 0.793 & Large \\
     Small World Coeff. & 84.1 (141.7) & 14.9 (17.0) & $<0.001$ & 0.919 & Large \\
    Coreness Ratio & 0.070 (0.064) & 0.353 (0.210) & $<0.001$ & -0.928 & Large \\
    Robustness Ratio & 0.326 (0.231) & 0.776 (0.215) & $<0.001$ & -0.833 & Large \\
    Avg. Constraint & 0.827 (0.081) & 0.918 (0.113) & $<0.001$ & -0.557 & Large \\
    Avg. Effective Size & 1.484 (0.271) & 1.185 (0.214) & $<0.001$ & 0.700 & Large \\
    \bottomrule
  \end{tabular}
  \caption*{\footnotesize \emph{Note}: All $p$-values except Collaboration Rate are significant after Bonferroni correction. Effect sizes: $|\delta| < 0.147$ = negligible, $0.147 \leq |\delta| < 0.33$ = small, $0.33 \leq |\delta| < 0.474$ = medium, $|\delta| \geq 0.474$ = large.}
\end{table}

\subsection{Separating Size Effects from Popularity Effects}

\begin{figure}[htbp]
    \centering
    \includegraphics[width=\textwidth]{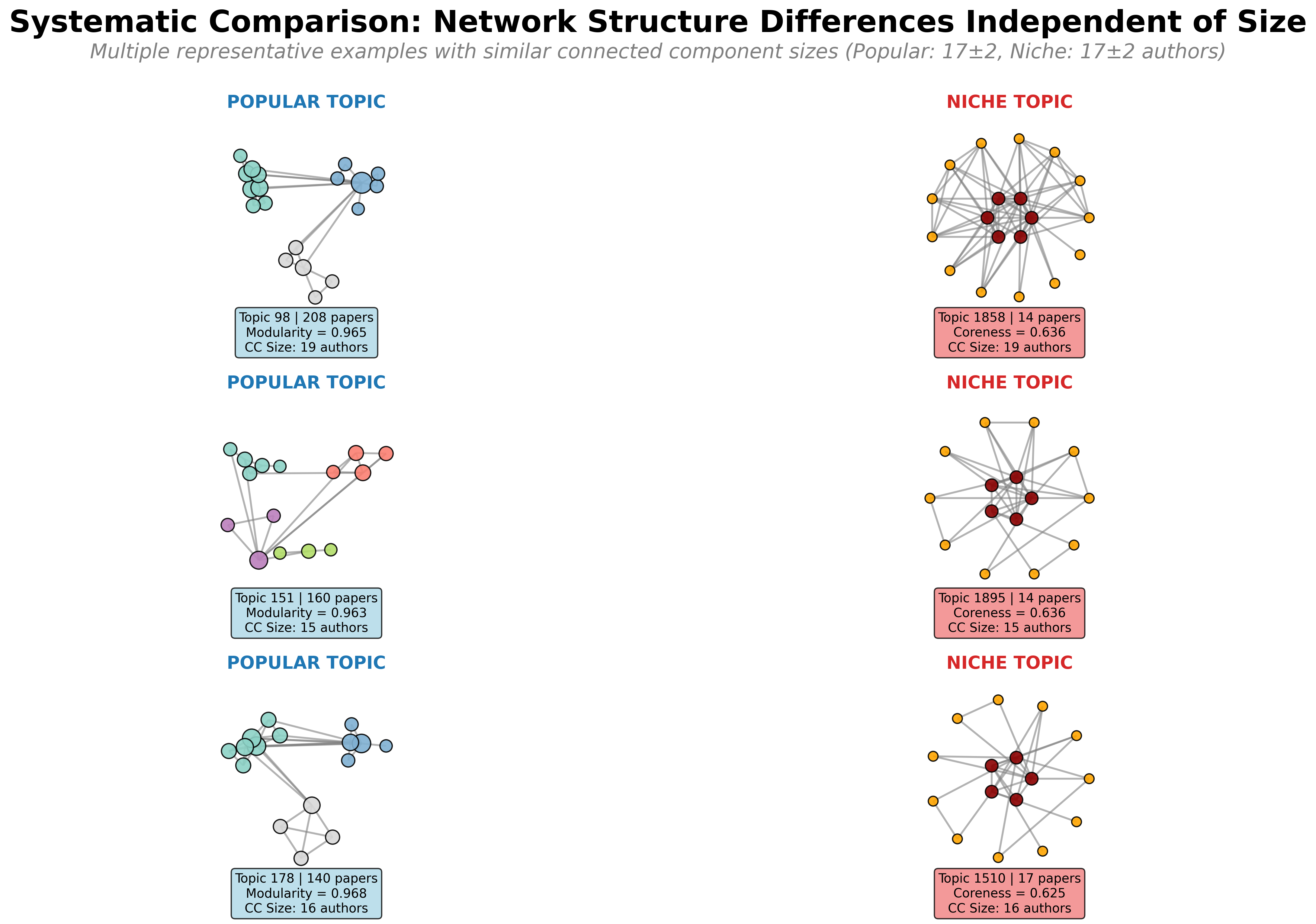} % Replace with your final figure file
    \caption{
         Collaboration networks for 3 exemplar pairs of popular and niche mathematical research topics, systematically selected to have similar connected component sizes. Each row shows one matched pair. (\textbf{Left}) Popular topics consistently exhibit modular community structure with distinct research groups (different colors), reflecting high modularity values. (\textbf{Right}) Niche topics consistently show core-periphery organization with central expert nodes (dark red, larger) densely connected to peripheral researchers (orange, smaller), reflecting high coreness ratios.}
    \label{fig:network_comparison}
\end{figure}

Network size correlations revealed substantial confounding for multiple metrics (Table \ref{tab:size_correlations}). The strongest associations appeared for robustness ratio ($r = -0.79$), coreness ratio ($r = -0.73$), and small-world coefficient ($r = 0.72$), confirming that a simple bivariate comparison is insufficient to isolate the effects of popularity. To visually illustrate our core finding, Figure \ref{fig:network_comparison} displays exemplar networks for a popular and a niche topic with identical author counts, highlighting the stark structural differences that persist even when size is held constant.

\begin{table}[htbp]
  \centering
  \caption{Network Size Correlations and Confounding Strength}
  \label{tab:size_correlations}
  \small
  \begin{tabular}{@{}lc@{}}
    \toprule
    Metric & Correlation with Size \\
    \midrule
    Robustness Ratio & -0.785*** \\
    Coreness Ratio & -0.733*** \\
    Small World Coefficient & 0.722*** \\
    Avg. Effective Size & 0.640*** \\
    Modularity & 0.616*** \\
    Avg. Constraint & -0.541*** \\
    Degree Centralization & -0.443*** \\
    Degree Assortativity & -0.323*** \\
    Repeated Collab. Rate & 0.242*** \\
    Collaboration Rate & 0.236*** \\
    \bottomrule
  \end{tabular}
  \caption*{\footnotesize \emph{Note}: Pearson correlation between metric and log-transformed author count. ***$p$ < 0.001.}
\end{table}

Regression analysis controlling for network size revealed three distinct patterns in how popularity effects manifest (Table \ref{tab:regression_results}).

\subsubsection{Robust Popularity Effects}
First, a core set of three metrics demonstrated robust popularity effects that persisted after size control. Modularity retained a substantial positive association with popularity ($\beta = 0.55, p < 0.001$), though its effect magnitude decreased by 56\% from the simple model. Coreness ratio maintained a strong negative relationship ($\beta = -0.73, p < 0.001$), preserving 46\% of its original effect size. Finally, average constraint exhibited an intriguing reversal; the effect was negative in the simple model but became strongly positive after controlling for size ($\beta = 0.46, p = 0.004$). These findings suggest that community fragmentation, expert-led hierarchies, and researcher constraint are genuine social adaptations to a topic's popularity.

\subsubsection{Size-Confounded and Marginal Effects}
Second, a majority of metrics (six of ten) revealed effects that were either fully confounded by size or only marginally significant. For metrics like robustness ratio and degree assortativity, the initial strong associations with popularity vanished entirely after controlling for network size ($p > 0.6$), indicating these are clear artifacts of scale. 

More subtly, degree centralization ($p=0.008$) and small-world coefficient ($p=0.026$) showed $p$-values that, while not surviving our stringent Bonferroni correction, could be considered marginally significant. This suggests the possibility of weaker, secondary popularity effects on these properties, which may become more apparent with even larger datasets or in different scientific domains. For the conservative purposes of our main analysis, however, we classify these as primarily scale-driven phenomena.

\subsubsection{Emergent Effects Through Statistical Suppression}
Third, collaboration rate displayed a striking emergent pattern. While the simple comparison showed no significant difference between groups, controlling for network size revealed a strong negative association ($\beta = -2.33, p < 0.001$). This apparent contradiction reflects a \textit{masking phenomenon} where a positive correlation between size and collaboration masks a negative relationship between popularity and collaboration. At equivalent network sizes, popular topics actually have lower collaboration rates, a relationship only visible after disentangling the two effects.

\begin{table}[htbp]
  \centering
  \captionsetup{justification=centering}
  \caption{Regression Results: Popularity Effects With and Without Size Control}
  \label{tab:regression_results}
  \scriptsize
  \begin{tabular}{l r@{ } c r@{ } c c c}
    \toprule
    & \multicolumn{1}{c}{\textbf{Simple Model}} & & \multicolumn{1}{c}{\textbf{Size Control Model}} & & & \\
    \cmidrule(lr){2-2} \cmidrule(lr){4-4}
    Metric & \multicolumn{1}{c}{$\beta$} & & \multicolumn{1}{c}{$\beta$} & & \multicolumn{1}{c}{Classification} & \multicolumn{1}{c}{R² Improv.} \\
    \midrule
    \multicolumn{7}{l}{\textbf{Robust Popularity Effects}} \\
    Modularity          & $1.26^{***}$ & & $0.55^{***}$ & & Robust & 2.4\% \\
    Coreness Ratio      & $-1.60^{***}$ & & $-0.73^{***}$ & & Robust & 3.2\% \\
    Avg. Constraint     & $-0.96^{***}$ & & $0.46^{**}$ & & Robust (Reversed) & 9.1\% \\
    \addlinespace
    \multicolumn{7}{l}{\textbf{Size-Confounded Effects}} \\
    Degree Centralization & $-0.88^{***}$ & & $-0.42$ & & Confounded & 1.1\% \\
    Small World Coeff.  & $1.75^{***}$ & & $-0.41^{*}$ & & Confounded & 10.1\% \\
    Robustness Ratio    & $-1.62^{***}$ & & $-0.05$ & & Confounded & 9.9\% \\
    Degree Assortativity& $-0.59^{***}$ & & $-0.003$ & & Confounded & 1.7\% \\
    Avg. Effective Size & $1.25^{***}$ & & $-0.28$ & & Confounded & 9.4\% \\
    Repeated Collab. Rate & $0.40^{***}$ & & $-0.23$ & & Confounded & 2.1\% \\
    \addlinespace
    \multicolumn{7}{l}{\textbf{Emergent Effect}} \\
    Collaboration Rate  & $0.05$ & & $-2.33^{***}$ & & Emergent & 27.9\% \\
    \bottomrule
  \end{tabular}
  \caption*{\footnotesize \emph{Note}: $\beta$ = standardized coefficient. Significance levels based on Bonferroni-corrected threshold ($\alpha=0.005$): \textsuperscript{***}$p<$0.001, \textsuperscript{**}$p<$0.005. \textsuperscript{*}$p<$0.05 is shown for reference but not considered robust. $R^2$ Improv. shows the increase in adjusted $R^2$ after adding the size control. All variables standardized before analysis.}
\end{table}

The explanatory power gained from including network size varied substantially, with model $R^2$ improvements ranging from a modest 1.1\% for degree centralization to a substantial 27.9\% for collaboration rate. Interaction models generally provided minimal additional explanatory power, suggesting that the relationships are primarily additive rather than multiplicative in nature. Taken together, these patterns reveal that collaboration networks are governed by dual mechanisms: universal scaling laws that create predictable size effects, and field-specific social dynamics that operate independently of—and sometimes in opposition to—these scaling relationships.

\section{Discussion}\label{sec:discussion}

\subsection{Collaboration Environments Across Mathematical Fields}\label{subsec:collab_environs}

Our analysis of 121,391 mathematical papers shows that research topics have fundamentally different collaboration environments, and these differences persist even after controlling for network size. This provides an empirical basis for understanding how topic selection shapes the context of a research career.

The distinction between the organization of popular and niche topics reflects different modes of knowledge production. Popular topics, which have significantly higher modularity (87.9\% vs. 68.8\%, $\delta$ = 0.79), are fragmented into distinct research communities. This structure is consistent with the classic concept of "invisible colleges" \citep{deSollaPrice1966, Crane1972} and is identified in modern network science by optimizing the modularity metric, $Q$ \citep{newman2006modularity, Fortunato2010}. The high modularity we observe ($\beta$ = 0.55 after size control) suggests that as fields grow, they subdivide to manage an expanding literature. For a researcher, this implies navigating a landscape of semi-autonomous communities, each with its own technical vocabulary and methods.

Conversely, niche topics maintain a pronounced core-periphery structure (7.0\% vs 35.3\%, $\delta = -0.93$). The high coreness ratio in these fields indicates a clear division between a dense core of central authors and a sparser periphery \citep{Borgatti2000, zhang2015}. This hierarchical organization is supported by high degree assortativity \citep{Newman2002}, where established experts preferentially connect with each other while maintaining supervisory links to peripheral members. This structure facilitates the efficient transmission of specialized knowledge when the pool of experts is small. For an early-career researcher, this places a premium on advisor selection and institutional placement to gain access to the field's core.

Furthermore, the emergent property whereby popular topics show lower collaboration rates at equivalent network sizes ($\beta = -2.33$, $p < 0.001$) adds another dimension to these considerations. This suggests that beyond topological differences, popular and niche fields may also cultivate different collaboration norms. Popular fields, with their established research programs and larger communities, might place a greater emphasis on individual contributions, whereas the tighter-knit, exploratory nature of niche fields may necessitate more interdependent teamwork.

\begin{comment}
The robustness of these findings is supported by our temporal sensitivity analysis (Appendix \ref{sec:appendix_tables}, Table \ref{tab:covid_sensitivity}), which confirms that these organizational patterns persisted through the disruption of the COVID-19 pandemic. While the pandemic temporarily affected some hierarchical metrics (degree centralization and constraint became non-significant during 2020-2021), the modular-hierarchical distinction remained stable. This consistency suggests these structures reflect stable organizational responses to different research contexts rather than temporary phenomena.
\end{comment}

\subsection{Brokerage Opportunities and Career Stage in Modular Fields}\label{subsec:brokerage_opportunities}

Our analysis of structural constraint reveals a complex relationship with popularity. While a simple bivariate comparison shows researchers in popular topics face less constraint, this relationship reverses after controlling for network size; they instead face significantly higher structural constraints ($\beta$ = 0.46, $p < 0.004$). This result extends Burt's theory of structural holes by incorporating a temporal, career-stage dimension \citep{Burt2004}.

The modular organization of popular fields creates numerous structural holes between communities, a configuration Burt identifies as advantageous for innovation. However, our data shows that the researchers occupying these brokerage positions are predominantly those with higher publication counts and seniority. This suggests a ``brokerage ladder'': a career progression from a constrained position within a module toward a boundary-spanning role.

This progression can be understood as a network-based mechanism for the ``Matthew Effect'' \citep{Merton1968}, a dynamic formalized in network science as ``preferential attachment'' \citep{Barabasi1999}. This principle posits that new collaborations are more likely to form with already-prominent researchers. Thus, when a valuable brokerage opportunity arises, it is the senior, high-status researchers who are most likely to attract that connection. The social capital required to span these boundaries is therefore a resource that accumulates over a career.

As \citet{everett2020unpacking} note, network size is an intrinsic part of the constraint measure. By statistically controlling for size, our analysis isolates the effect of structural position relative to a researcher's peers. The high average constraint in popular fields is therefore not paradoxical; it reflects the typical early-career experience. A researcher must first embed within a specific community to build reputation, with brokerage opportunities emerging only as their career matures.

\subsection{Implications for Mathematical Career Navigation}\label{subsec:implications_career}

These structural differences carry practical implications for researchers at various career stages, though we emphasize that our analysis characterizes collaboration environments rather than predicting career outcomes. The Math Research Compass operationalizes these insights by providing transparent access to collaboration structures across 1,938 identified topics.

For early-career mathematicians, our findings suggest that topic selection involves an implicit choice between contrasting collaborative environments. Those drawn to popular areas should anticipate working within modular communities where establishing position within a specific research cluster becomes paramount. Success in such environments likely requires different strategies than in niche fields, where direct access to central experts provides clearer mentorship pathways but potentially fewer alternative routes should initial connections prove unproductive.

The emergent property whereby popular topics show lower collaboration rates at equivalent network sizes ($\beta = -2.33$, $p < 0.001$) adds another dimension to these considerations. This suggests that beyond structural differences, popular and niche fields may cultivate different collaboration norms, with popular fields potentially emphasizing individual contributions within established research programs.

\subsection{Methodological Contributions and Limitations}\label{subsec:strengths_limitations}

A key methodological contribution of this work is our two-stage analytical approach, which rigorously separates universal scaling effects from genuine popularity-driven organizational patterns. By first comparing groups and then using regression to control for network size, we demonstrate how univariate analyses can be misleading. The dramatic attenuation or reversal of several effects after size control (Table \ref{tab:regression_results}) underscores the necessity of this method for future studies of scientific collaboration.

While our findings are robust within this framework, several limitations define the scope of our conclusions. Our reliance on arXiv data, while comprehensive for many theoretical fields, may incompletely represent mathematical collaboration in applied areas where conference proceedings or industry partnerships are more common. Furthermore, our cross-sectional design establishes strong associations but precludes direct causal inference; future longitudinal work is needed to track how network structures evolve as topics gain or lose popularity. Finally, this study characterizes collaborative environments but does not link them to career outcomes like academic placement or grant success, which remains a valuable avenue for subsequent research.

We proactively addressed two other potential methodological concerns. First, while our author name disambiguation (AND) pipeline achieves a high validated precision of 86.0\%, we acknowledge that residual errors could influence network metrics. False negatives (missed merges), the more likely error type for our conservative algorithm, would tend to make networks appear more fragmented, suggesting our reported differences in modularity and coreness may be \textit{underestimates} of the true effect. 

Second, to ensure our findings were not an artifact of the COVID-19 pandemic, we performed a full temporal sensitivity analysis, splitting our dataset into ``peak pandemic'' (2020--2021) and ``post-peak'' (2022--2025) periods. As detailed in Appendix \ref{sec:appendix_tables} (Table \ref{tab:covid_sensitivity}), the modular-hierarchical duality remained stable and highly significant across both eras. This provides strong evidence that the observed structures are fundamental features of mathematical collaboration, not transient phenomena. Interestingly, the only metrics that showed instability (degree centralization and average constraint) were not significant during the pandemic, suggesting that some hierarchical features may have been temporarily flattened by the global shift to remote work.

Future work should build on this foundation by applying our validated pipeline to longitudinal data and linking structural patterns to concrete career outcomes. Examining how researchers successfully transition between communities within modular fields could also provide actionable insights for those seeking to expand their collaborative reach.

\subsection{Conclusion}\label{subsec:conclusion}

Our comprehensive analysis reveals that mathematical collaboration networks exhibit predictable structural variations based on topic popularity, with these differences persisting after controlling for network size. The duality between universal scaling laws and field-specific organizational patterns suggests that while some network properties emerge mechanically with growth, others reflect genuine adaptations to different research contexts.

By documenting these patterns systematically and making them accessible through the Math Research Compass, we aim to democratize knowledge about collaboration structures that previously remained tacit within mathematical communities. While we stop short of prescriptive career advice, we believe that transparency about these structural realities enables more informed decision-making as researchers navigate the mathematical landscape. Understanding whether one is entering a modular, school-based field or a hierarchical, expert-centered domain represents valuable context for calibrating expectations and developing appropriate collaborative strategies.

%\appendix
\begin{appendices}

\section{Computational Details for Network Metrics}\label{sec:appendix_metrics}

All metrics were computed using Python's NetworkX library (v3.4.2) \citep{Hagberg2008}. Key implementation details are noted below.

\subsection{Repeated Collaboration Rate}
This metric was calculated as the ratio of unique author pairs with more than one joint publication to the total number of unique collaborating pairs:
\[ \text{Repeated Collab. Rate} = \frac{\text{Number of author pairs with edge weight } > 1}{\text{Total number of unique author pairs with an edge}} \]

\subsection{Small-World Coefficient ($\omega$)}
To ensure a robust calculation for potentially disconnected networks, our custom implementation first identified the LCC of each topic network. The actual clustering coefficient $(C)$ and average shortest path length ($L$) were computed on this LCC. The expected values for an equivalent Erdős-Rényi random graph ($C_{\text{rand}}$ and $L_{\text{rand}}$) were then calculated based on the number of nodes and edges of the LCC, providing a consistent null model. For networks with over 200 nodes, path lengths were estimated using random sampling to ensure computational feasibility.

\subsection{Robustness Ratio}
This metric was calculated by simulating and comparing two node removal scenarios. We measured the size of the LCC after removing the top 10\% of nodes by degree (targeted attack) and compared it to the LCC size after removing 10\% of nodes selected uniformly at random (random failure). The ratio of these two resulting LCC sizes quantifies the network's vulnerability to targeted attacks.

\subsection{Note on Small Networks}
Networks with fewer than 3 nodes typically returned default values (e.g., 0 or NaN depending on the metric) or were handled to ensure computational stability; such cases were minimal.

\section{Author Name Disambiguation Pipeline}\label{sec:AND_pipeline}

To ensure the construction of high-precision collaboration networks, we developed and implemented a conservative, multi-stage author name disambiguation (AND) pipeline named \texttt{AuthorDisambiguator}. The pipeline is designed to accurately merge common name variations (e.g., `J. Doe' and `John Doe') while aggressively preventing the incorrect merging of distinct authors with similar names (e.g., `Pengcheng Xie' and `Pengxu Xie'), a common failure mode in less conservative systems. The process is organized into three main stages, preceded by a normalization step.

\subsection{Data Parsing and Normalization}

For each of the 121,391 papers, author lists were parsed from the raw ArXiv metadata format. All unique author strings were then subjected to a rigorous normalization procedure to create a consistent representation for comparison:
\begin{enumerate}
    \item \textbf{Lowercase Conversion:} All strings were converted to lowercase.
    \item \textbf{Diacritic Removal:} Diacritics were removed by normalizing to the NFD Unicode form and encoding to ASCII (e.g., ``Fran\c{c}ois'' becomes ``francois'').
    \item \textbf{Particle Standardization:} Common name particles (e.g., 'van', 'der', 'de', 'von') were standardized to prevent them from being treated as separate name components during splitting.
    \item \textbf{Punctuation and Character Removal:} All punctuation (except for internal hyphens and apostrophes, which are handled by later tokenization) and non-alphanumeric characters were removed.
\end{enumerate}

\subsection{Stage 1: Initial Similarity Graph Construction}

To efficiently identify potential merge candidates, we first constructed an undirected similarity graph where nodes represent unique normalized author strings.
\begin{description}
    \item[\textbf{Blocking:}] To ensure computational feasibility across $\approx$120,000 unique names, we employed a blocking strategy. Name pairs were only considered for similarity comparison if they shared the same last name.
    
    \item[\textbf{Edge Creation:}] An edge was added between two nodes (authors) in the graph if they satisfied strict string similarity criteria, defined as either:
    \begin{enumerate}
        \item \textbf{Initial Expansion:} One name is a clear initialism of the other (e.g., ``J. Doe'' and ``John Doe'').
        \item \textbf{High String Similarity:} The names have a Levenshtein distance ratio greater than 0.95, as calculated by Python's \texttt{difflib.SequenceMatcher}.
    \end{enumerate}
\end{description}

\subsection{Stage 2: Cluster Merging}

The connected components of the similarity graph, representing clusters of potentially identical authors, were then analyzed using a set of conservative, rule-based heuristics to decide whether a merge was safe.
\begin{description}
    \item[\textbf{Conservative Handling of Large Clusters:}] Clusters containing more than 50 name variants were flagged for cautious treatment. The algorithm did not attempt to merge the entire cluster, but instead sought to identify and merge only highly confident subgroups within it, leaving ambiguous cases unresolved. This prevents cascading errors in clusters corresponding to common last names.
    
    \item[\textbf{First Name Compatibility Logic:}] The core of the merging process relied on assessing the compatibility of full first names within a cluster. A cluster was approved for merging only if all extracted first names were determined to be plausible variants of each other. This was assessed using the following rules:
    \begin{enumerate}
        \item \textbf{Abbreviation/Diminutive Check:} One name is a standard abbreviation or diminutive of another. This includes initialism matching (e.g., `Rob' and `Robert') and a pre-defined dictionary of common Western variants (e.g., `William'/`Bill',`Christopher'/`Chris').
        \item \textbf{Context-Sensitive String Similarity:} For names not covered by the above, a high string similarity was required. Crucially, to handle the ambiguity common in Romanized East Asian names, we applied a stricter similarity threshold. Based on a list of common pinyin syllables (e.g., `wei', `ming', `xiao', `zhang'), if a name pair was identified as potentially Asian, a similarity threshold of 0.92 was required; otherwise, a threshold of 0.87 was used.
    \end{enumerate}
A cluster was only merged into a single canonical profile if all its first-name variants were compatible under this logic. For ambiguous clusters containing multiple distinct and incompatible first names, the algorithm attempted to partition the cluster into smaller, internally consistent subgroups for merging.
\end{description}

\subsection{Stage 3: Network-Based Co-occurrence Merging}

The final stage used co-authorship and publication patterns to resolve remaining ambiguities among canonical profiles that were not merged by string similarity alone.
\begin{description}
    \item[\textbf{Blocking:}] Candidate pairs for this stage were limited to authors who shared the same last name and the same first initial.
    
    \item[\textbf{Weighted Jaccard Similarity:}] Each candidate pair was evaluated using a weighted Jaccard similarity score based on their sets of co-authors and their sets of publications:
    \begin{itemize}
        \item Let $C_1, C_2$ be the sets of co-authors for authors 1 and 2.
        \item Let $P_1, P_2$ be the sets of papers for authors 1 and 2.
        \item The combined similarity was calculated as:
        \begin{equation}
            S = w \cdot \frac{|C_1 \cap C_2|}{|C_1 \cup C_2|} + (1-w) \cdot \frac{|P_1 \cap P_2|}{|P_1 \cup P_2|}
        \end{equation}
    \end{itemize}
    
    \item[\textbf{Thresholding:}] A pair was merged if $S > 0.5$ and each author had a minimum of 2 papers in the dataset. The co-author weight was set to $w = 0.6$ to prioritize shared collaborative circles, a strong signal of identity.
\end{description}

\subsection{Final Canonical Name Resolution}

After all merging stages, the resulting author-to-canonical-name mapping was resolved transitively to ensure that all names in a merge chain pointed to a single, final canonical profile. The representative name for each merged cluster was chosen as the most complete name variant available, prioritizing names with more components and fewer initials. This process reduced the 120,767 unique raw author strings to 117,883 canonical author profiles, with a manually validated precision of 86.0\% on a random sample of 100 merges.

\section{Supplemental Tables}\label{sec:appendix_tables}

\begin{table}[htbp]
  \centering
  \captionsetup{justification=centering}
  \caption{Full Regression Models with Interaction Effects}
  \label{tab:appendix_interaction}
  %\scriptsize
  \begin{tabular}{l c c c c}
    \toprule
    & \multicolumn{3}{c}{\textbf{Standardized Coefficients ($\beta$) from Interaction Model}} & \\
    \cmidrule(lr){2-4}
    Metric & \textbf{Popularity} & \textbf{log(Authors)} & \textbf{Interaction Term} & \textbf{Adj. R² (Full Model)} \\
    \midrule
    Modularity              & $0.208$        & $0.811^{***}$   & $0.655$         & 0.407 \\
    Coreness Ratio          & $-0.507^{**}$  & $-0.700^{***}$  & $-0.354^{**}$   & 0.562 \\
    Avg. Constraint         & $0.641^{***}$  & $-0.865^{***}$  & $-0.395$        & 0.306 \\
    Collaboration Rate      & $-3.654^{***}$ & $3.116^{***}$   & $1.383^{***}$   & 0.614 \\
    \midrule
    Degree Centralization   & $-0.656^{***}$ & $0.215$         & $0.435$         & 0.215 \\
    Small World Coeff.      & $-0.218$       & $0.387^{**}$    & $0.347$         & 0.539 \\
    Robustness Ratio        & $-0.087$       & $-0.560^{***}$  & $0.063$         & 0.617 \\
    Degree Assortativity    & $0.074$        & $-0.356^{*}$    & $-0.280$        & 0.105 \\
    Avg. Effective Size     & $-0.270$       & $0.593^{***}$   & $0.021$         & 0.412 \\
    Repeated Collab. Rate   & $-0.545^{**}$  & $0.764^{***}$   & $0.579$         & 0.081 \\
    \bottomrule
  \end{tabular}
  \caption*{\footnotesize \emph{Note}: Full interaction models (Metric ~ Popularity + log(Authors) + Interaction). A significant `Interaction Term` coefficient indicates that the relationship between size and the metric differs for popular vs. niche topics. Significance levels: \textsuperscript{***}$p<$0.001, \textsuperscript{**}$p<$0.01, \textsuperscript{*}$p<$0.05.}
\end{table}

% In your preamble, make sure you have: \usepackage{booktabs}, \usepackage{siunitx}

\begin{table}[htbp]
  \centering
  \caption{Robustness of Effect Sizes (Cliff's Delta) Across Different Topic Model Granularities}
  \label{tab:robustness_effect_sizes}
  % \scriptsize
  \sisetup{
      table-format = -1.3, % Format for numbers: sign, one digit before, three after decimal
      table-align-text-post = false % Don't align text like 'negligible'
  }
  \small % Use a slightly smaller font for the table
  \begin{tabular}{@{} l S S S S @{}}
    \toprule
    & \multicolumn{4}{c}{\textbf{Effect Size (Cliff's $\delta$) at different \texttt{min\_topic\_size} settings}} \\
    \cmidrule(l){2-5}
    \textbf{Network Metric} & {\textbf{10}} & {\textbf{15 (main)}} & {\textbf{20}} & {\textbf{25}} \\
    \midrule
    \multicolumn{5}{l}{\textit{\textbf{Metrics where Popular > Niche}}} \\
    Modularity              & 0.905 & 0.785 & 0.794 & 0.846 \\
                            & \scriptsize{[0.88, 0.93]} & \scriptsize{[0.74, 0.83]} & \scriptsize{[0.73, 0.85]} & \scriptsize{[0.80, 0.89]} \\
    \addlinespace
    Avg. Effective Size     & 0.818 & 0.705 & 0.727 & 0.693 \\
                            & \scriptsize{[0.78, 0.85]} & \scriptsize{[0.65, 0.76]} & \scriptsize{[0.66, 0.79]} & \scriptsize{[0.63, 0.76]} \\
    \addlinespace
    Repeated Collab. Rate   & 0.590 & 0.284 & 0.308 & 0.345 \\
                            & \scriptsize{[0.52, 0.65]} & \scriptsize{[0.20, 0.37]} & \scriptsize{[0.20, 0.42]} & \scriptsize{[0.25, 0.44]} \\
    \midrule
    \multicolumn{5}{l}{\textit{\textbf{Metrics where Niche > Popular}}} \\
    Coreness Ratio          & -0.678 & -0.929 & -0.928 & -0.902 \\
                            & \scriptsize{[-0.74, -0.61]} & \scriptsize{[-0.95, -0.90]} & \scriptsize{[-0.96, -0.89]} & \scriptsize{[-0.94, -0.86]} \\
    \addlinespace
    Robustness Ratio        & -0.894 & -0.840 & -0.906 & -0.866 \\
                            & \scriptsize{[-0.92, -0.86]} & \scriptsize{[-0.88, -0.80]} & \scriptsize{[-0.94, -0.87]} & \scriptsize{[-0.91, -0.82]} \\
    \addlinespace
    Avg. Constraint         & -0.361 & -0.565 & -0.557 & -0.512 \\
                            & \scriptsize{[-0.43, -0.29]} & \scriptsize{[-0.63, -0.50]} & \scriptsize{[-0.64, -0.47]} & \scriptsize{[-0.59, -0.43]} \\
    \addlinespace
    Degree Assortativity    & -0.310 & -0.326 & -0.348 & -0.278 \\
                            & \scriptsize{[-0.40, -0.22]} & \scriptsize{[-0.40, -0.24]} & \scriptsize{[-0.45, -0.24]} & \scriptsize{[-0.38, -0.18]} \\
    \addlinespace
    Degree Centralization   & -0.075 & -0.558 & -0.658 & -0.594 \\
                            & \scriptsize{[-0.15, 0.00]} & \scriptsize{[-0.62, -0.49]} & \scriptsize{[-0.73, -0.58]} & \scriptsize{[-0.67, -0.51]} \\
    \midrule
    \multicolumn{5}{l}{\textit{\textbf{Metric with No Significant Difference}}} \\
    Collaboration Rate      & -0.055 & -0.027 & 0.135 & 0.028 \\
                            & \scriptsize{[-0.13, 0.02]} & \scriptsize{[-0.11, 0.05]} & \scriptsize{[-0.03, 0.24]} & \scriptsize{[-0.07, 0.13]} \\
    \bottomrule
  \end{tabular}
  \caption*{\footnotesize \emph{Note}: This table shows the stability of the observed effect sizes (Cliff's $\delta$) across four different topic models generated with varying \texttt{min\_topic\_size} parameters (10, 15, 20, 25). The first value in each cell is the observed Cliff's $\delta$, and the values in brackets represent the 95\% bootstrap confidence interval (10,000 iterations). The direction and magnitude of the significant effects remain highly consistent across all specifications, demonstrating the robustness of our core findings to the granularity of the topic model.}
\end{table}

\begin{table}[htbp]
  \centering
  \captionsetup{justification=centering}
  \caption{Performance Comparison of Binary vs. Continuous Popularity Measures}
  \label{tab:supp_binary_vs_continuous}
  \footnotesize % Use small font for a dense table
  \sisetup{
    table-format = -1.3,
  } % Setup for number alignment and significance stars

  \begin{tabular}{
    l
    c
    S[table-format=1.3]
    c 
    c
    S[table-format=1.3]
    c 
    c
  }
    \toprule
    & \multicolumn{2}{c}{\textbf{Binary Model}} & & \multicolumn{2}{c}{\textbf{Best Continuous Model}} & & \\
    \cmidrule(lr){2-3} \cmidrule(lr){5-6}
    {Network Metric} & {\textbf{$\beta$}} & {\textbf{Adj. R²}} & {\textbf{Type}} & {\textbf{$\beta$}} & {\textbf{Adj. R²}} & {\textbf{Cont. Better?}} & {\textbf{Effect Consist.}} \\
    \midrule
    Collaboration Rate        & -3.915$^{***}$ & 0.335 & {Log Quadratic} & -4.886$^{***}$ & 0.974 & {Yes} & {Medium (62\%)} \\
    Repeated Collab. Rate     & -0.200          & 0.068 & {Log Quadratic} & -0.631$^{***}$ & 0.076 & {Yes} & {Medium (62\%)} \\
    Degree Centralization     & -1.182$^{**}$  & 0.313 & {Log Quadratic} & -0.809$^{***}$ & 0.224 & {No}  & {\textbf{High (85\%)}} \\
    Degree Assortativity      & -0.028          & 0.146 & {Log Quadratic} & 0.300           & 0.110 & {No}  & {Medium (46\%)} \\
    Modularity                & 1.102$^{**}$   & 0.496 & {Log Quadratic} & 0.061           & 0.399 & {No}  & {\textbf{High (85\%)}} \\
    Small World Coeff.        & -3.403$^{***}$ & 0.675 & {Z-score + Size} & 0.908$^{***}$ & 0.642 & {No}  & {Medium (69\%)} \\
    Coreness Ratio            & -1.020$^{***}$ & 0.636 & {Log Quadratic} & -0.196          & 0.554 & {No}  & {\textbf{High (85\%)}} \\
    Robustness Ratio          & -0.345          & 0.765 & {Robust + Size} & 0.055           & 0.604 & {No}  & {Medium (46\%)} \\
    Avg. Constraint           & 0.932$^{**}$   & 0.394 & {Log Quadratic} & 1.000$^{***}$  & 0.331 & {No}  & {\textbf{High (100\%)}} \\
    Avg. Effective Size       & -0.743$^{*}$   & 0.538 & {Log Quadratic} & -0.431$^{**}$  & 0.406 & {No}  & {\textbf{High (85\%)}} \\
    \bottomrule
  \end{tabular}
  \caption*{\footnotesize \emph{Note}: All models control for network size. This table compares the binary classification model against the best-performing of six continuous popularity measures for each metric. \textbf{$\beta$} = Standardized regression coefficient. \textbf{Adj. $R^2$} = Adjusted $R$-squared. \textbf{Effect Consist.} = Percentage of models where the popularity term was significant ($p$<0.05) in the same direction. Significance levels: \textsuperscript{***}$p<$0.001, \textsuperscript{**}$p<$0.01, \textsuperscript{*}$p<$0.05.}
\end{table}

\begin{table}[htbp]
  \centering
  \caption{Robustness of Baseline Findings to COVID-19 Temporal Effects}
  \label{tab:covid_sensitivity}
  % Setup siunitx for number alignment and to handle +/- signs
  \sisetup{
      table-format = -1.3, 
      table-space-text-post = \textsuperscript{***}
  }
  \small % Use a slightly smaller font
  \begin{tabular}{@{} l c c c @{}}
    \toprule
    & \multicolumn{1}{c}{\textbf{2020--2021}} & \multicolumn{1}{c}{\textbf{2022--2025}} & \\
    \cmidrule(lr){2-2} \cmidrule(lr){3-3}
    \textbf{Network Metric} & {\textbf{Cliff's $\delta$}} & {\textbf{Cliff's $\delta$}} & \textbf{Consistent?}\textsuperscript{a} \\
    \midrule
    \multicolumn{4}{l}{\textit{Consistent Effects (Popular > Niche)}} \\
    Modularity              & 0.959\textsuperscript{***} & 0.875\textsuperscript{***} & \textbf{Yes} \\
    Avg. Effective Size     & 0.751\textsuperscript{***} & 0.706\textsuperscript{***} & \textbf{Yes} \\
    Small World Coefficient & 0.794\textsuperscript{***} & 0.857\textsuperscript{***} & \textbf{Yes} \\
    \addlinespace
    \multicolumn{4}{l}{\textit{Consistent Effects (Niche > Popular)}} \\
    Coreness Ratio          & -0.585\textsuperscript{***} & -0.929\textsuperscript{***} & \textbf{Yes} \\
    Robustness Ratio        & -0.715\textsuperscript{***} & -0.800\textsuperscript{***} & \textbf{Yes} \\
    Degree Assortativity    & -0.431\textsuperscript{***} & -0.419\textsuperscript{***} & \textbf{Yes} \\
    \addlinespace
    \multicolumn{4}{l}{\textit{Inconsistent or Non-Significant Effects}} \\
    Degree Centralization   & 0.005 & -0.542\textsuperscript{***} & No\textsuperscript{b} \\
    Avg. Constraint         & -0.127 & -0.467\textsuperscript{***} & No\textsuperscript{c} \\
    \bottomrule
  \end{tabular}
  \caption*{\footnotesize \emph{Note}: Comparison of effect sizes for key metrics across two temporal periods. \textsuperscript{a}Consistency requires the effect to be statistically significant and in the same direction in both periods. \textsuperscript{b,c}The effect was not statistically significant during the 2020--2021 period. Significance levels based on Bonferroni-corrected $p$-values: \textsuperscript{***}$p<$0.005.}
\end{table}

\begin{table}[ht]
\centering
\caption{Manual Validation of Representative BERTopic-Identified Topics}
\label{tab:manual_BERT}
\begin{tabular}{lccccl}
\hline
\textbf{Topic} & \textbf{Size} & \textbf{Quality} & \textbf{Coherence} & \textbf{Artifact?} & \textbf{Mathematical Area} \\
\textbf{Keywords (truncated)} & \textbf{(papers)} & \textbf{(1-5)} & \textbf{(1-5)} & & \\
\hline
\multicolumn{6}{l}{\textit{Small Topics (15-25 papers)}} \\
Fermion systems, causal... & 15 & 4 & 4 & No & Mathematical Physics \\
Cherednik algebras, rational... & 16 & 5 & 5 & No & Representation Theory \\
Parking functions, parking... & 17 & 4 & 4 & No & Combinatorics \\
Vortex filaments, filament... & 18 & 4 & 4 & No & Fluid Dynamics \\
\hline
\multicolumn{6}{l}{\textit{Medium Topics (50-100 papers)}} \\
Ricci curvature, Ricci flat... & 54 & 5 & 5 & No & Differential Geometry \\
Iwasawa theory, Iwasawa main... & 58 & 5 & 5 & No & Number Theory \\
Weil-Petersson metric... & 68 & 5 & 5 & No & Teichmüller Theory \\
Siegel modular forms... & 88 & 5 & 5 & No & Number Theory \\
\hline
\multicolumn{6}{l}{\textit{Large Topics (150+ papers)}} \\
Hopf algebras, Hopf algebra... & 171 & 5 & 5 & No & Quantum Groups \\
Mean curvature flow... & 155 & 5 & 5 & No & Geometric Analysis \\
Bergman spaces, Bergman... & 206 & 5 & 5 & No & Complex Analysis \\
\hline
\multicolumn{6}{l}{\textit{Very Large Topics (400+ papers)}} \\
Elliptic curves, elliptic... & 659 & 5 & 5 & No & Algebraic Geometry \\
Knot invariants, knot... & 698 & 5 & 5 & No & Knot Theory \\
\hline
\end{tabular}
\label{tab:topic_validation}
\caption*{
\small
 \emph{Note}: Quality and Coherence rated 1-5 by expert assessment. Sample shows 13 of 100 manually validated topics. Full validation: 74\% high quality (4-5), 26\% medium quality (3), 0\% low quality or artifacts.
}
\end{table}

\end{appendices}
%\newpage
\clearpage
\bibliography{references}
\end{document}